\documentclass[12pt,preprint]{aastex}

\usepackage[center]{caption}

\shorttitle{KP Cyg}
\shortauthors{Andrievsky et al.}

\begin{document}

\title{KP Cyg: an Unusual Metal-rich RR Lyr Type Star of Long Period}

\author{ S.M. Andrievsky\altaffilmark{1} and V.V. Kovtyukh}
\affil{Department of Astronomy and Astronomical Observatory, Odessa 
National University, T.G. Shevchenko Park, 65014, Odessa, Ukraine}
\email{scan@deneb1.odessa.ua, val@deneb1.odessa.ua}

\author{George Wallerstein}
\affil{ Department of Astronomy, University of Washington, Seattle, WA 98195} 
\email{wall@astro.washington.edu}

\author{S.A. Korotin}
\affil{Department of Astronomy and Astronomical Observatory, Odessa 
National University, T.G. Shevchenko Park, 65014, Odessa, Ukraine}
\email{serkor@skyline.od.ua}

\author{Wenjin Huang}
\affil{ Department of Astronomy, University of Washington, Seattle, WA 98195} 
\email{hwenjin@astro.washington.edu}

\begin{abstract}

We present the results of a detailed spectroscopic study of the long period ($P=0.856$ days)
RR~Lyrae star, KP~Cyg.  We derived abundances of many chemical elements including the
light species, iron-group elements and elements of the s-processes. Most RR~Lyrae
stars with periods longer than 0.7 days are metal-deficient objects.  Surprisingly,
our results show that KP~Cyg is very metal rich ([Fe/H] $= +0.18\pm 0.23$).
By comparison with a number of short period ($P=1\sim 6$ days), metal-rich CWB stars, we
suggest that KP~Cyg may be a very short period CWB star (BL~Her star) rather than an RR~Lyrae 
star.  As seen in some CWB stars, KP~Cyg shows strong excesses of carbon and nitrogen
in its atmosphere.  This indicates that the surface of KP~Cyg has been polluted by
material that has undergone helium burning (to enhance carbon) and proton capture
(to transform carbon into nitrogen).  We also note that UY CrB, whose period is 0.929
days, also shows an enhancement of C and N, and that two carbon cepheids of short 
period, V553 Cen and RT TrA, show similar excesses of carbon and nitrogen.
 
\end{abstract}


\section{Introduction}

The RR~Lyrae stars have been known for a long time to be an old population
with diverse metallicity (from near solar to [Fe/H]$\sim -2.5$).
They are present in most globular clusters but in widely different
numbers ranging from zero to about 200 (Clement et al. 2001).
The general properties of RR~Lyrae stars can be found in detail in the book
by Smith (1995).  Most RR~Lyrae stars in the field and in clusters are readily
divided into two major categories: the RRc-type with periods from approximately
0.20 d to 0.45 d, and the RRab type with periods from about 0.4 to 1.0 days.
There is a statistical correlation between metallicities and periods of RR~Lyrae
stars that the stars with longer periods are generally more metal-poor.

Very few RR~Lyrae stars are known with periods longer than 0.75 days.  In his 
classic paper on the metallicity of RR~Lyrae stars Preston (1959) included 
a few long period stars that he found to be relatively metal-rich.  However 
he subsequently noted that the data of some of those stars, mostly their
periods, were in error.  Among the recently discovered RR~Lyrae stars
in the Northern Sky Variability Survey (NSVS), Kinemuchi et al.\ (2009)
have found 21 variables with periods between 0.74 and 0.86 days whose
metallicities range from [Fe/H]$= -1.50$ to $-2.15$ dex, similar to the
RR~Lyrae stars with periods of 0.60 to 0.75 days.

The general picture of the long-period RR~Lyrae stars having low metallicity
has not been strongly challenged until the discovery of some long-period
RR~Lyrae stars in the two globulars, NGC~6388 and 6441 (Pritzl et al. 2000).
These two massive clusters have unusual color-magnitude diagrams.  Their red
giant branches are relatively faint indicating that they are relatively
metal-rich for globular clusters.  A spectroscopic analysis of red giants in
NGC 6388 showed that [Fe/H] = $-0.7$ (Wallerstein, Kovtyukh, and Andrievsky
2007).  For NGC 6441, Gratton et al.\ (2007) found a metallicity of [Fe/H] =
$-0.34$.  In addition, for NGC 6441, Clementini et al.\ (2005) found a metallicity
of $-0.7$ from a sample of its RR Lyrae stars.  Both horizontal branches
consist of a well populated red clump accompanied by a significant number
of blue stars including a blue tail in NGC 6441.  From Preston's correlation
of period and metallicity such high metallicties indicate that the RR Lyrae
stars should have short periods.  However Pritzl et al. found a wide range
of periods in these clusters including numerous variables with periods larger
than 0.75 days.

In an attempt to find field RR~Lyrae stars that are similar to
those with long periods in NGC~6388 and 6441, we have observed a few RR~Lyraes
with periods greater than 0.75 days.  Among them, KP~Cyg appears to be the most
unusual.  First noted by Vogt (1970) and subsequently confirmed by others
(Loomis et al. 1988, Schmidt 2002), KP~Cyg has a period of about 0.856 days,
quite long for an RR~Lyrae star.  Preston (1959) found its $\Delta~S$ value to
be around 0, implying that KP~Cyg is a metal-rich object.  Other than this
information, we know little about the star.  In this paper we present the
results of an analysis of the chemical composition in this star's atmosphere.

\section{Observation and Data Reduction}

Spectra of KP~Cyg were obtained using the echelle spectrograph on
the 3.5-m telescope, at the Apache Point Observatory (APO).  The usable
wavelength coverage, limited by the red-sensitive $2048 \times 2048$ CCD chip,
runs from 4000 \AA~ to 9000 \AA.  The resolving power is about 35000.  The integration
time of each exposure was set to 20$\sim$30 minutes to reach good S/N ratios, and to
avoid heavy contamination from cosmic ray events.  We then combined the spectra
that were taken sequentially within one hour, and measured the S/N ratio at the 
continuum level per pixel in the very clean region between 7550 and 7600\AA~of 
the combined spectra.

Table 1 lists the dates (JD), phases, the heliocentric radial velocities ($V_{\rm r}$), the
derived values of T$_{\rm eff}$ and $\log~g$ for each phase, and signal-to-noise
ratio of each spectrum.  Our $V_{\rm r}$ measurements confirm that the period of KP Cyg
is around 0.9 days.  We also note the presence of anomalous structures in H$\alpha$
such as line doubling and emission.  A comparison of the abundances derived from the
spectra having anomalous H$\alpha$ profiles with those derived from the spectra showing
normal H$\alpha$ profiles 
demonstrates that there is no very significant difference among the derived [Fe/H] values. 
We suspect that the anomalous features in the H$\alpha$ line profile originate from the
layers far above the line forming region that is relevant to our abundance analysis.

\begin{table*}
\scriptsize
\begin{center}
\caption[]{Observations of KP~Cyg and its atmospheric parameters}
\begin{tabular}{ccccccccc}
\hline
 Date   &  JD 2450000+    & phase & S/N & $V_{\rm r}$ (km s$^{-1}$)& T$_{\rm eff}$,K & $\log~g$ & 
V$_{\rm t}$ (km s$^{-1}$) & Remarks \\
\hline 
2005-11-13&  3687.606& .135 &  46 &\phn8.6   & 7050 & 2.6 & 2.5 &                   \\
2006-11-05&  4044.672& .299 &  47 &   13.1   & 6600 & 3.0 & 3.3 &                   \\
2007-05-01&  4221.963& .430 &  73 &   20.3   & 6450 & 3.4 & 3.5 & H$\alpha$ emission\\
2005-09-24&  3637.607& .720 &  88 &   35.6   & 6300 & 2.4 & 3.2 & H$\alpha$ emission\\
2006-10-03&  4011.720& .800 &  47 &   27.5   & 6650 & 3.0 & 4.2 & H$\alpha$ emission\\
2005-09-13&  3626.711& .991 &  91 & $-$1.8   & 7400 & 3.0 & 2.5 &                   \\
\hline
\end{tabular}
\end{center}
\end{table*}

The spectra were extracted from the raw frames using standard
IRAF\footnote{http://iraf.noao.edu} procedures. 
The continuum level placement, wavelength calibration and equivalent widths 
measurements were performed with DECH20 code (Galazutdinov 1992).  The final
equivalent width measurements are presented in Table A1, whose full content
is available only in the electronic edition.  A small portion of the table
is given here to illustrate its format and content.  The equivalent widths
of some very strong or seriously blended lines were not measured.  Instead,
for these lines, we compared their NLTE synthesized profiles directly with
the observed spectra in the abundance analysis (see \S3.2).

\section{Method of Analysis}

\subsection{Atmospheric Parameters and LTE Elemental Abundances}

The LTE elemental abundances were derived using the Kurucz's WIDTH9 (Kurucz 1996)
code with the model atmospheres interpolated from the ATLAS9 model grid. We used 
$\log~gf$ values derived from an inverted solar analysis (Kovtyukh \& Andrievsky 1999).

The atmospheric parameters of KP~Cyg (T$_{\rm eff}$, $\log~g$, V$_{\rm t}$)
were determined by meeting a few requirements in our spectroscopic analysis: 
T$_{\rm eff}$ and V$_{\rm t}$ were constrained by minimizing the dependence of
the derived iron abundance from each Fe~I line on their excitation potentials
and  equivalent widths; $\log~g$ was determined by requiring an ionization
equilibrium between Fe~I and Fe~II. While $\log~g$ may be calculated using
the known luminosities and masses of RR~Lyrae stars, we preferred to use the
method based on the FeI/FeII ionization equilibrium because the acceleration
of the atmosphere during its pulsation cycle is then automatically taken into
account.

\subsection{NLTE Elemental Abundances}

The NLTE effects were considered in deriving the abundances of such elements as carbon,
nitrogen, oxygen, sodium, magnesium, aluminum, sulfur, potassium, strontium, and barium.
The details of the atomic models used for calculating the NLTE line profiles are described 
in a series of papers by Andrievsky et al. (2001, 2007, 2008, 2009, 2010a,b) for carbon,
sodium, aluminum, barium, magnesium, potassium, and strontium respectively,
Korotin (2009) --- for sulfur, Mishenina et al. (2000) --- for oxygen,
Lyubimkov et al. (2010) --- for nitrogen.  The NLTE abundances of these elements were 
derived using a NLTE spectrum synthesis code adapted from MULTI (see Carlsson (1986) and  
Korotin et al. (1999a,b)).  The list of the lines used in the NLTE calculations and their 
parameters are given in Table~2.

\begin{table*}
\scriptsize
\begin{center}
\caption[]{List of the lines used for NLTE calculations}
\begin{tabular}{rcccc|rcccc}
\hline 
 Ion  & $\lambda$ & $\chi$(eV) & $\log~gf$ &$\log~\epsilon$(El)$_{\odot}$ & Ion  & $\lambda$ & $\chi$
(eV) & $\log~gf$ & $\log~\epsilon$(El)$_{\odot}$\\
\hline
 CI   &  5052.17 & 7.685&  $-1.30$  &  8.43 & NaI  &  4982.81 & 2.104& $-0.96$ &  6.25 \\
      &  5380.34 & 7.685&  $-1.62$  &       &      &  4982.81 & 2.104& $-1.91$ &       \\
      &  6014.83 & 8.643&  $-1.58$  &       &      &  5682.63 & 2.102& $-0.70$ &       \\
      &  6413.55 & 8.771&  $-2.00$  &       &      &  5688.19 & 2.104& $-1.39$ &       \\
      &  6587.61 & 8.537&  $-1.13$  &       &      &  5688.20 & 2.104& $-0.46$ &       \\
      &  6655.51 & 8.537&  $-1.79$  &       &      &  6154.22 & 2.102& $-1.53$ &       \\
      &  7087.83 & 8.647&  $-1.44$  &       &      &  6160.74 & 2.104& $-1.23$ &       \\
      &  7100.12 & 8.643&  $-1.47$  &       &      &  8183.25 & 2.102&  ~0.26  &       \\
      &  7108.93 & 8.640&  $-1.59$  &       &      &  8194.79 & 2.104& $-0.44$ &       \\
      &  7111.47 & 8.640&  $-1.09$  &       &      &  8194.82 & 2.104&  ~0.51  &       \\
      &  7113.17 & 8.647&  $-0.77$  &       & MgI  &  4167.27 & 4.346& $-0.77$ &  7.58 \\
      &  7115.17 & 8.643&  $-0.93$  &       &      &  4702.99 & 4.346& $-0.52$ &       \\
      &  7115.18 & 8.640&  $-1.47$  &       &      &  5172.68 & 2.712& $-0.38$ &       \\
      &  7116.99 & 8.647&  $-0.91$  &       &      &  5183.60 & 2.717& $-0.16$ &       \\
      &  7119.65 & 8.643&  $-1.15$  &       &      &  5528.40 & 4.346& $-0.62$ &       \\
      &  7473.30 & 8.771&  $-2.04$  &       &      &  5711.09 & 4.346& $-1.72$ &       \\
      &  7476.17 & 8.771&  $-1.57$  &       & AlI  &  6696.02 & 3.143& $-1.48$ &  6.43 \\
      &  7483.44 & 8.771&  $-1.37$  &       &      &  6698.66 & 3.143& $-1.78$ &       \\
      &  7848.24 & 8.848&  $-1.73$  &       &      &  7835.30 & 4.022& $-0.64$ &       \\
      &  7852.86 & 8.851&  $-1.68$  &       &      &  7836.13 & 4.022& $-1.64$ &       \\
      &  7860.88 & 8.851&  $-1.15$  &       &      &  7836.13 & 4.022& $-0.45$ &       \\
      &  8335.14 & 7.685&  $-0.44$  &       &      &  8772.87 & 4.022& $-0.24$ &       \\
 NI   &  7442.29 &10.330&  $-0.39$  &  7.89 &      &  8773.90 & 4.022& $-0.02$ &       \\
      &  7468.31 &10.336&  $-0.19$  &       &      &  8773.90 & 4.022& $-1.32$ &       \\
      &  8184.86 &10.330&  $-0.28$  &       & SI   &  6538.60 & 8.046& $-0.93$ &  7.16 \\
      &  8188.01 &10.326&  $-0.29$  &       &      &  6743.44 & 7.866& $-1.20$ &       \\
      &  8210.71 &10.330&  $-0.73$  &       &      &  6743.53 & 7.866& $-0.85$ &       \\
      &  8216.33 &10.336&   ~0.09   &       &      &  6743.64 & 7.866& $-0.95$ &       \\
      &  8223.12 &10.330&  $-0.32$  &       &      &  6748.57 & 7.868& $-1.32$ &       \\
      &  8242.38 &10.336&  $-0.32$  &       &      &  6748.68 & 7.868& $-0.73$ &       \\
      &  8680.28 &10.336&   ~0.35   &       &      &  6748.84 & 7.868& $-0.53$ &       \\
      &  8683.40 &10.330&   ~0.09   &       &      &  6756.85 & 7.870& $-1.67$ &       \\
      &  8686.14 &10.326&  $-0.30$  &       &      &  6757.01 & 7.870& $-0.83$ &       \\
      &  8703.24 &10.326&  $-0.32$  &       &      &  6757.17 & 7.870& $-0.24$ &       \\
      &  8711.70 &10.330&  $-0.23$  &       &      &  8693.93 & 7.870& $-0.51$ &       \\
      &  8718.83 &10.336&  $-0.33$  &       &      &  8694.63 & 7.870&  ~0.08  &       \\
\end{tabular}
\end{center}
\end{table*}

\begin{table*}
\tablenum{2}
\scriptsize
\begin{center}
\caption[]{List of the lines used for NLTE calculations (continued)}
\begin{tabular}{rcccc|rcccc}
\hline
 Ion  & $\lambda$ & $\chi$(eV) & $\log~gf$ &$\log~\epsilon$(El)$_{\odot}$ & Ion  & $\lambda$ & $\chi$
(eV) & $\log~gf$ & $\log~\epsilon$(El)$_{\odot}$\\
\hline
 OI   &  5577.34 & 1.967&  $-8.24$  &  8.71 & KI   &  7664.91 & 0.000& $ 0.13$ &  5.11 \\
      &  6155.97 &10.740&  $-0.67$  &       &      &  7698.97 & 0.000& $-0.17$ &       \\
      &  6156.76 &10.741&  $-0.45$  &       & SrII &  4077.71 & 0.000&  ~0.15  &  2.92 \\
      &  6158.17 &10.741&  $-0.31$  &       &      &  4161.79 & 2.940& $-0.41$ &       \\
      &  6300.30 & 0.000&  $-9.75$  &       &      &  4215.52 & 0.000& $-0.18$ &       \\
      &  6363.77 & 0.020&  $-10.30$ &       & BaII &  4554.03 & 0.000&  ~0.08  &  2.17 \\
      &  7771.94 & 9.146&   ~0.33   &       &      &  4554.05 & 0.000& $-0.79$ &       \\
      &  7774.16 & 9.146&   ~0.19   &       &      &  4554.00 & 0.000& $-1.01$ &       \\
      &  7775.38 & 9.146&  $-0.03$  &       &      &  5853.68 & 0.604& $-1.00$ &       \\
      &  8446.24 & 9.521&  $-0.52$  &       &      &  6141.71 & 0.704& $-0.08$ &       \\
      &  8446.35 & 9.521&   ~0.18   &       &      &  6496.89 & 0.604& $-0.46$ &       \\
      &  8446.75 & 9.521&  $-0.05$  &       &      &  6496.89 & 0.604& $-1.32$ &       \\
      &          &      &           &       &      &  6496.90 & 0.604& $-1.55$ &       \\
\hline
\end{tabular}
\end{center}
\end{table*}

Because some of the NLTE-treated lines are blended with other lines, we used the combined NLTE 
and LTE synthetic spectra when we compared the NLTE line profiles with the observed spectra.  
This was done using the code SYNTHV (Tsymbal 1996), which was developed for LTE spectrum synthesis.  
We calculated the synthetic spectra for the selected regions comprising the lines of interest and 
all nearby lines that are found in VALData-base (Kupka et al. 2000).  Meanwhile, for each line 
treated in NLTE, the corresponding $b$-factors (the ratios of the NLTE to LTE populations in 
the involved energy levels) were calculated by MULTI and then fed into SYNTHV for calculating 
the NLTE line source functions.  Some examples of the NLTE profile fitting in KP~Cyg spectra are 
shown in Fig.~1.

\begin{figure}
\resizebox{12.5cm}{!}{\includegraphics{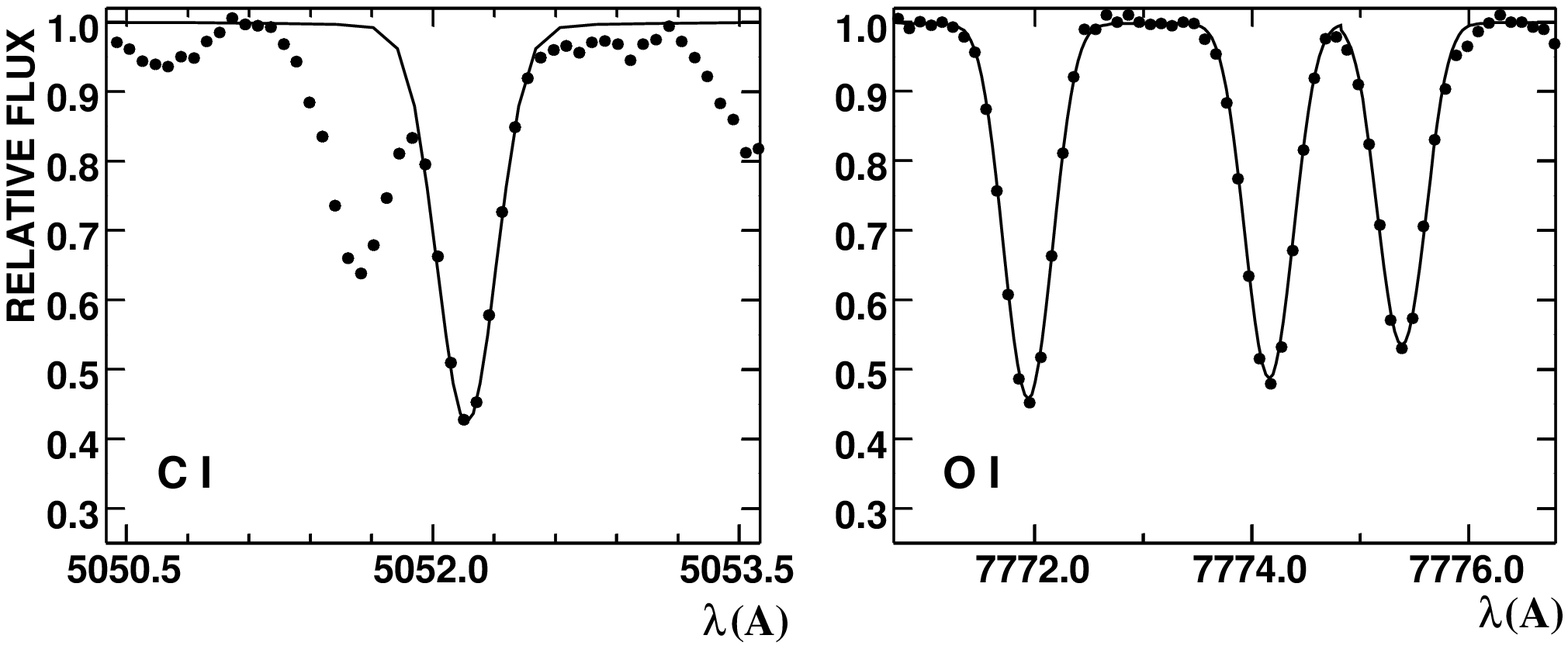}}
\resizebox{12.5cm}{!}{\includegraphics{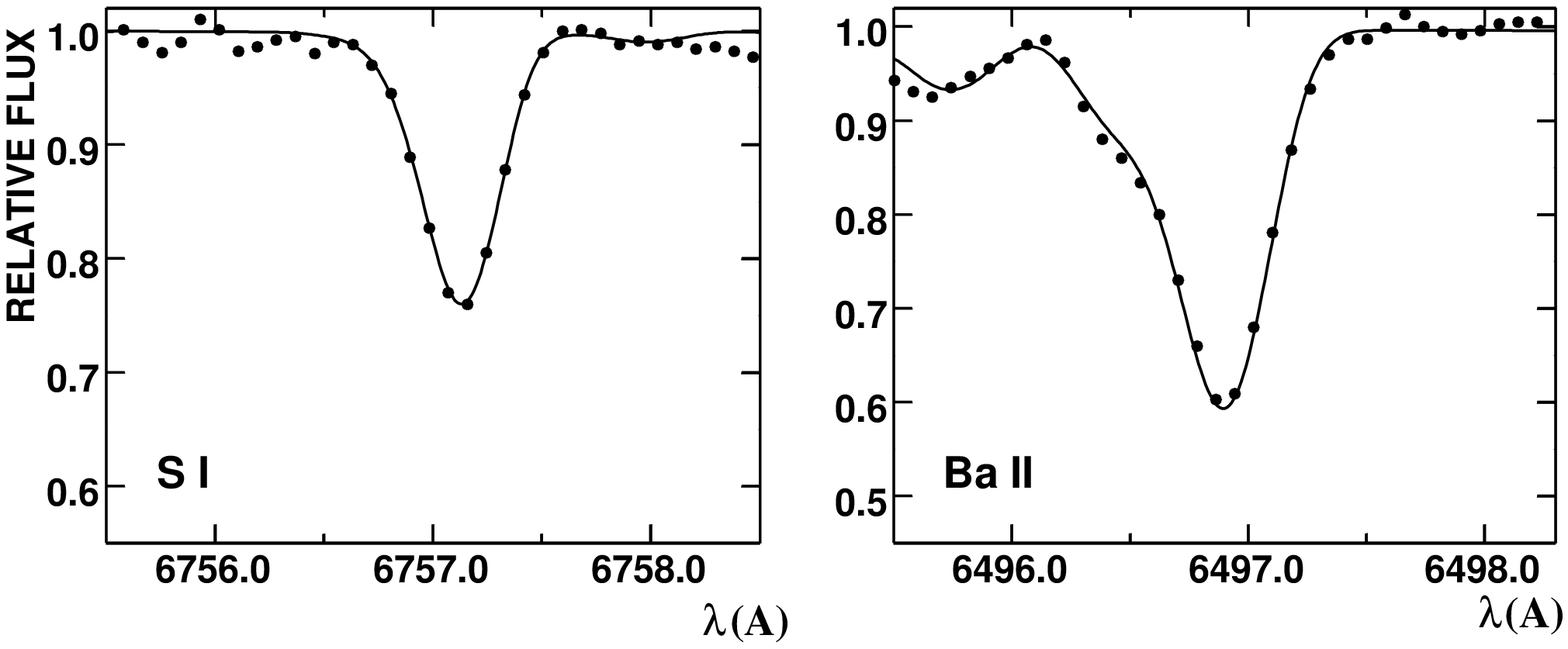}}
\caption[]{Examples of the best-fit synthesized NLTE profiles ({\it solid lines})
are over-plotted on the observed spectra ({\it filled dots}).}
\end{figure}

 Our results of the NLTE abundance analysis of KP~Cyg need to be compared to the sun. 
Hence we chose to use the same source for the model atmospheres for both the Sun and KP~Cyg.  
We used the Kurucz's solar model to derive our NLTE abundances in the Sun in the 
same way as we did for KP~Cyg. Solar equivalent widths were obtained from observations
of the asteroid, Vesta, obtained  
by Don York, using the same instrument (the APO echelle spectrograph) as was
used for the KP~Cyg spectra. The derived NLTE abundances of the Sun (based on the line list of 
Table~2) are given in the last column of Table~2.

In Table~3, we show the sensitivity of the NLTE oxygen abundance to the variation of the adopted 
atmospheric parameters.  As one can see, the changes of the parameters within our observational 
uncertainty ($\Delta$~T$_{\rm eff} \approx \pm 150$K, $\Delta~\log~g \approx \pm 0.2$, and
$\Delta$~V$_{\rm t} \approx \pm 0.3$ km~s$^{-1}$) cause only small fluctuations in the derived 
oxygen abundance ($< 0.2$ dex).

\begin{table}
\begin{center}
\caption[]{Parameter variation and NLTE oxygen abundance for KP Cyg}
\begin{tabular}{cccc}
\hline
T$_{\rm eff}$ &  $\log~g$ & V$_{\rm t}$&$\log~\epsilon$(O) \\
\hline
 6600 & 3.0  & 3.3 &  9.05 \\
 6450 & 3.0  & 3.3 &  9.15 \\
 6750 & 3.0  & 3.3 &  8.96 \\
 6600 & 3.2  & 3.3 &  9.15 \\
 6600 & 2.8  & 3.3 &  8.98 \\
 6600 & 3.0  & 3.6 &  8.96 \\
 6600 & 3.0  & 3.0 &  9.14 \\
\hline
\end{tabular}
\end{center}
\end{table}

\begin{table*}
\scriptsize
\begin{center}
\caption[]{Elemental abundances in KP Cyg}
\begin{tabular}{rrrrrrrrrrrrr}
\hline
\hline
\multicolumn{1}{c}{}&\multicolumn{3}{c}{$\phi=0.135$}&
\multicolumn{3}{c}{$\phi=0.299$}&\multicolumn{3}{c}{$\phi=0.430$}&
\multicolumn{3}{c}{$\phi=0.720$}\\
\hline
\multicolumn{1}{c}{}&\multicolumn{3}{c}{7050/2.6/2.5}&
\multicolumn{3}{c}{6600/3.0/3.3}&\multicolumn{3}{c}{6450/3.4/3.5}&
\multicolumn{3}{c}{6300/2.4/2.3}\\
\hline
Ion & [El/H] & $\sigma$ & N &[El/H] & $\sigma$ & N &[El/H] &
$\sigma$ & N &[El/H]&$\sigma$&N\\
\hline
 CI* & +0.86 & 0.13 & 23 & +0.92 & 0.13 & 18 & +0.95 & 0.11 & 19 & +0.77 & 0.15 &17\\
 NI* & +0.91 & 0.14 & 13 & +1.18 & 0.14 & 11 & +1.32 & 0.10 & 11 & +0.91 & 0.15 & 6\\
 OI* & +0.41 & 0.12 &  4 & +0.29 & 0.12 &  6 & +0.34 & 0.12 &  6 & +0.24 & 0.12 & 3\\
NaI* & +0.54 & 0.12 &  6 & +0.45 & 0.12 &  5 & +0.48 & 0.10 &  6 & +0.50 & 0.10 & 6\\
MgI* & +0.14 & 0.12 &  6 &--0.05 & 0.12 &  3 & +0.07 & 0.10 &  5 &--0.05 & 0.12 & 5\\
AlI* & +0.47 & 0.17 &  2 & +0.37 & 0.13 &  3 & +0.44 & 0.15 &  6 & +0.42 & 0.15 & 4\\
SiI  & +0.16 & 0.12 & 10 & +0.32 & 0.06 & 14 & +0.29 & 0.10 & 17 & +0.14 & 0.18 & 7\\
SiII & +0.11 &  --  &  1 &   --  &  --  & -- & +0.25 &  --  &  2 & +0.12 &  --  & 2\\
 SI* & +0.26 & 0.13 &  5 & +0.32 & 0.13 &  4 & +0.34 & 0.11 &  6 & +0.21 & 0.12 & 2\\
 KI* & +0.27 & 0.12 &  1 &--0.20 & 0.12 &  1 &   --  &   -- & -- &--0.10 & 0.12 & 1\\
 CaI & +0.04 & 0.19 &  9 & +0.03 & 0.12 &  9 & +0.03 & 0.15 &  8 & +0.19 & 0.23 & 8\\
ScII &--0.11 & 0.07 &  4 & +0.32 & 0.17 &  6 & +0.37 & 0.16 &  5 &--0.07 & 0.18 & 5\\
 TiI &--0.20 &  --  &  2 & +0.15 & 0.12 &  4 & +0.11 & 0.13 &  6 & +0.08 & 0.20 & 3\\
TiII &--0.20 & 0.04 &  3 & +0.13 & 0.12 &  5 & +0.14 & 0.14 &  5 &--0.08 & 0.16 & 3\\
 VII & +0.04 &  --  &  2 & +0.21 &  --  &  1 & +0.42 & 0.10 &  3 & +0.36 &  --  & 1\\
 CrI &--0.07 & 0.07 &  3 & +0.09 & 0.14 &  4 & +0.02 & 0.12 &  6 &--0.15 & 0.19 & 3\\
CrII &--0.05 & 0.08 &  5 & +0.17 & 0.08 &  5 & +0.14 & 0.11 &  3 &--0.04 & 0.18 & 4\\
 MnI & +0.11 &  --  &  2 & +0.15 & 0.05 &  2 & +0.16 & 0.19 &  3 & +0.34 &  --  & 2\\
 FeI & +0.09 & 0.12 & 68 & +0.29 & 0.10 & 72 & +0.30 & 0.15 & 88 & +0.05 & 0.19 &59\\
FeII & +0.09 & 0.13 & 17 & +0.29 & 0.08 & 15 & +0.31 & 0.12 & 18 & +0.08 & 0.19 &10\\
 NiI & +0.06 & 0.13 & 15 & +0.36 & 0.13 & 23 & +0.16 & 0.08 & 25 & +0.16 & 0.15 &12\\
SrII*&  --   &  --  & -- &  --   &  --  & -- & +0.12 & 0.17 &  2 &    -- &  --  &--\\
 YII & +0.08 & 0.21 &  3 & +0.35 & 0.16 &  4 & +0.25 & 0.15 &  5 & +0.08 &      & 2\\
BaII*& +0.20 & 0.14 &  4 &--0.15 & 0.10 &  4 &--0.20 & 0.10 &  4 &--0.35 & 0.10 & 3\\
NdII &    -- &  --  &  - & +0.12 & 0.02 &  2 & +0.24 & 0.20 &  3 & +0.25 &  --  & 2\\
\hline
\end{tabular}
\end{center}

\end{table*}

\begin{table*}
\tablenum{4}
\scriptsize
\begin{center}
\caption[]{Elemental abundances in KP Cyg (continued)}
\begin{tabular}{rrrrrrrr}
\hline
\hline
\multicolumn{1}{c}{}&\multicolumn{3}{c}{$\phi=0.800$}&
\multicolumn{3}{c}{$\phi=0.991$}&\multicolumn{1}{c}{}\\
\hline
\multicolumn{1}{c}{}&\multicolumn{3}{c}{6650/3.0/4.2}&
\multicolumn{3}{c}{7400/3.0/2.5}&\multicolumn{1}{c}{Mean}\\
\hline
Ion & [El/H] & $\sigma$ & N &[El/H] & $\sigma$ & N &[El/H]\\
\hline
  CI* &  +0.79 &  0.15 &  14 & +0.67 &  0.12 &  16 &   +0.83 \\ 
  NI* &  +1.14 &  0.15 &   9 & +1.09 &  0.12 &   8 &   +1.10 \\ 
  OI* &  +0.29 &  0.12 &   4 & +0.34 &  0.14 &   4 &   +0.32 \\ 
 NaI* &  +0.51 &  0.12 &   5 & +0.38 &  0.14 &   4 &   +0.48 \\ 
 MgI* & --0.13 &  0.12 &   6 &--0.13 &  0.15 &   4 &  --0.02 \\ 
 AlI* &  +0.68 &   --  &   1 &    -- &   --  &  -- &   +0.45 \\ 
 SiI  &  +0.34 &  0.11 &  20 & +0.16 &  0.18 &   8 &   +0.25 \\ 
SiII  &  +0.18 &   --  &   1 &   --  &   --  &  -- &   +0.17 \\ 
  SI* &  +0.27 &  0.13 &   2 & +0.14 &  0.14 &   3 &   +0.26 \\ 
  KI* & --0.10 &  0.12 &   1 & +0.10 &  0.12 &   1 &  --0.01 \\ 
 CaI  &  +0.05 &  0.14 &  10 &--0.15 &  0.12 &   7 &   +0.03 \\ 
ScII  &  +0.15 &  0.18 &   7 & +0.11 &  0.13 &   3 &   +0.14 \\ 
 TiI  &  +0.03 &  0.13 &   7 &   --  &   --  &  -- &   +0.05 \\ 
TiII  &  +0.04 &  0.06 &   7 & +0.03 &   --  &   1 &   +0.02 \\ 
 VII  &  +0.02 &  0.07 &   3 & +0.22 &   --  &   1 &   +0.20 \\ 
 CrI  &  +0.00 &  0.18 &   9 &   --  &   --  &  -- &  --0.01 \\ 
CrII  &  +0.05 &  0.10 &   8 & +0.14 &  0.14 &   3 &   +0.06 \\ 
 MnI  &  +0.15 &  0.17 &   6 &--0.14 &   --  &   1 &   +0.14 \\ 
 FeI  &  +0.18 &  0.12 & 137 & +0.12 &  0.18 &  55 &   +0.18 \\ 
FeII  &  +0.16 &  0.13 &  30 & +0.13 &  0.11 &  12 &   +0.18 \\ 
 NiI  &  +0.11 &  0.11 &  24 & +0.20 &  0.19 &   8 &   +0.18 \\ 
 SrII*&  +0.05 &  0.15 &   1 & +0.00 &  0.15 &   2 &   +0.07 \\ 
 YII  & --0.01 &  0.11 &   6 &--0.13 &   --  &   1 &   +0.13 \\ 
BaII* & --0.15 &  0.10 &   4 &--0.34 &  0.13 &   4 &  --0.16 \\ 
NdII  & --0.20 &   --  &   1 &  --   &   --  &   --&   +0.13 \\ 
\hline
\end{tabular}
\end{center}
* - NLTE abundances


\end{table*}


\section{Discussion}

The final derived elemental abundances are given in Table~4. The abundances of 
the iron-group elements in KP~Cyg are consistent with the results by Preston (1959) 
($\Delta$~S = 0). According to the combined data presented by Smith (1995, Fig. 3.9), 
there are no RR~Lyr stars in globular clusters both with metallicity higher than 
$-$1.0 and periods larger than 0.8 days except for a few stars in the two puzzling clusters, 
NGC~6388 and 6441, and a single star in 47~Tuc.  The super-solar metallicity and unusually 
long period makes KP~Cyg even more peculiar than the stars in NGC~6388 and 6441.
It casts a doubt on the RR~Lyrae classification of KP~Cyg.

RR~Lyrae stars and short period type II~Cepheids(CWB for short,
also called BL~Her stars) are known to have a helium-burning core and a
hydrogen-burning shell surrounding the core.  In the CN cycle, ${}^{12}$C
is converted to ${}^{14}$N through the reaction chain, 
${}^{12}$C($p$,$\gamma$)${}^{13}$N($\beta^+$,$\nu$)${}^{13}$C($p$,$\gamma$)${}^{14}$N.
In addition, ${}^{12}$C may capture an $\alpha$-particle to produce ${}^{16}$O.  
${}^{13}$C may become a source of neutrons through the ${}^{13}$C($\alpha$,n)${}^{16}$O 
reaction.  The neutrons then may be captured by nuclei of the iron peak elements to 
create Co and Cu, as well as many other heavier elements. Analysis of the abundances 
of CNO as well as light, and heavy s-process elements in these stars is the key 
to understanding how these processes function inside a star.

Our NLTE results for CNO clearly show that carbon and nitrogen are significantly
in excess in KP~Cyg, while oxygen demonstrates only a moderate overabundance.
An overabundance can be also noted for sodium and aluminum.  Altogether these facts
testify that this star has experienced dredge-up of the materials processed in CNO, 
NeNa (and perhaps MgAl) cycles.  Similar enhancements of C and N in some of the
short period type II cepheids, i.e. CWB type stars, were recently noted by
Maas, Giridhar, \& Lambert (2007). 
In Table~5 we show the values of $\log~\epsilon$~(C+N+O) and $\log~\epsilon$~(C+N)
for our program star and several short period ($P < 7$ days) CWB stars from
Maas et al. (2007).  All 7 stars are more metal-rich than the RR~Lyr stars in
globular clusters. The Sun is listed for comparison as well (the necessary input
CNO abundances are from Table~2).  

\begin{table*}
\begin{center}
\caption[]{A Comparison of the combined C, N, and O abundances in KP~Cyg
with short-period CWB stars (Maas et al. 2007), tow carbon cepheids, and the Sun.}
\begin{tabular}{ccrcc}
\hline
\hline
 Star  & P(days)   & [Fe/H]  & $\log~\epsilon$(C+N+O) & $\log~\epsilon$(C+N) \\
\hline
KP Cyg & 0.9     &   +0.18 & 9.59 & 9.45 \\
UY CrB & 0.9     & $-$0.40 & 9.20 & 8.94 \\
\hline
BX Del & 1.1     & $-$0.24 & 9.43 & 9.29 \\
VY Pyx & 1.2     & $-$0.46 & 9.18 & 8.97 \\
BL Her & 1.3     & $-$0.18 & 9.22 & 9.04 \\
SW Tau & 1.6     &   +0.18 & 9.73 & 9.66 \\
AU Peg & 2.4     & $-$0.24 & 9.11 & 8.80 \\
DQ And & 3.2     & $-$0.50 & 8.71 & 8.42 \\
TX Del & 6.2     &   +0.06 & 9.50 & 9.25 \\
\hline
V553 Cen & 2.1   &   +0.04 & 9.68 & 9.51 \\
RT TrA & 2.0     &   +0.41 & 9.68 & 9.61 \\
\hline
Sun    &\nodata  &   +0.00 & 8.93 & 8.54 \\
\hline
\end{tabular}
\end{center}
\end{table*}

\begin{figure}[h]
\resizebox{\hsize}{!}{\includegraphics{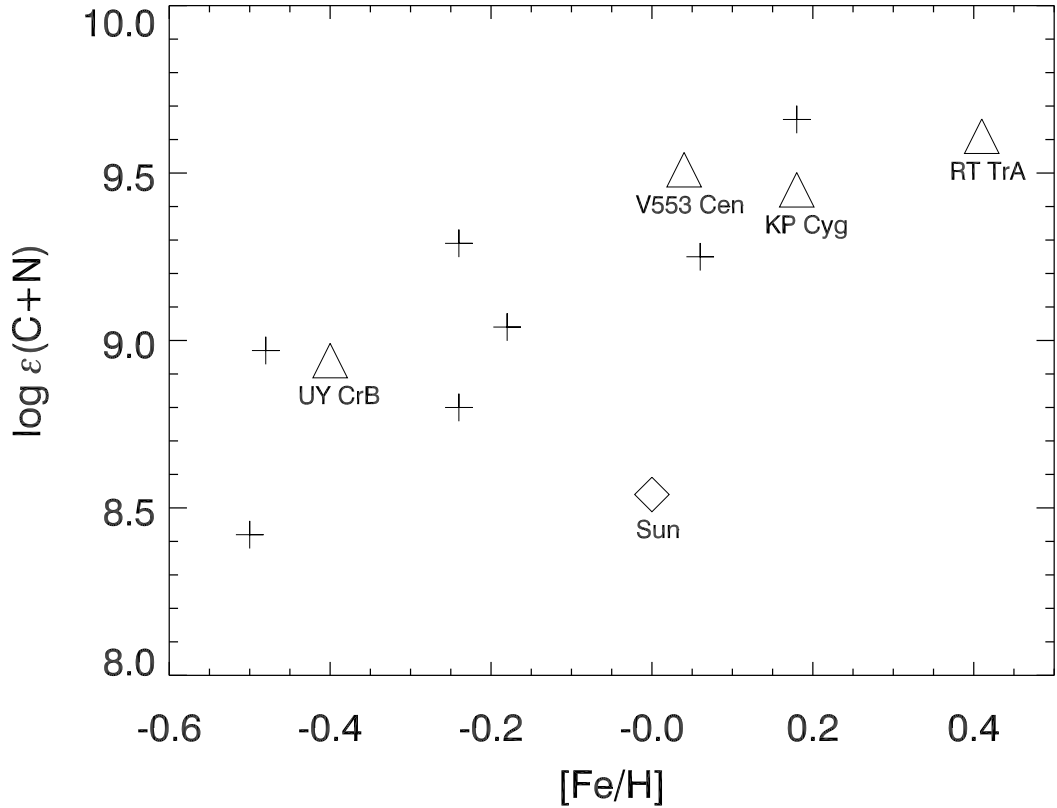}}
\caption[]{$\log~\epsilon$~(C+N) vs. [Fe/H] for KP~Cyg, 
UY~CrB, the short-period CWB stars ({\it as plus symbols})
from Maas et al. (2007), and the Sun.}
\label{Fig02}
\end{figure}

\begin{figure}[h]
\resizebox{\hsize}{!}{\includegraphics{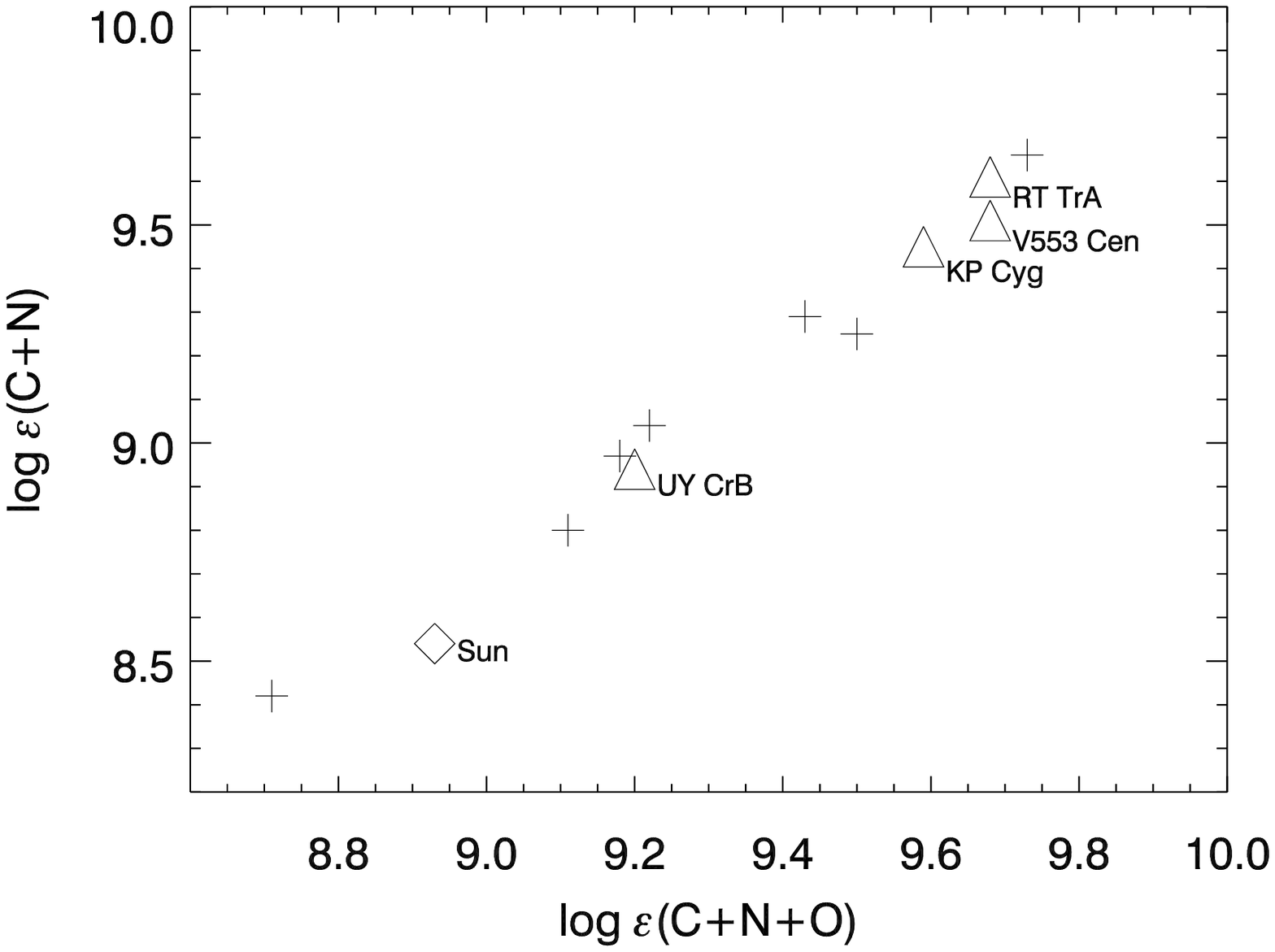}}
\caption[]{$\log~\epsilon$~(C+N) vs. $\log~\epsilon$~(C+N+O) for KP~Cyg, 
UY~CrB, the short-period CWB stars from Maas et al. (2007)({\it as plus
symbols}), and the Sun.}
\label{Fig03}
\end{figure}

Fig.~2 shows that the C and N excesses in the short period CWB stars are very similar 
to that of KP~Cyg.  As Maas et al. have suggested the excess of C must be due to helium 
burning during or after the core flash. The excess of N must be due to proton capture by C.  
The oxygen abundances do not show the great excesses of N. This means that O was not 
significantly enhanced by the capture of $\alpha$-particles by C. The Fig.~3 shows that 
the high values of $\log~\epsilon$~(C+N+O) is mainly due to the excess of C and N, while the
influence of O is very limited.  It is clear that the enhancement of N in these stars is 
due to the combination of the triple-alpha reaction and proton capture rather than just the 
re-arrangement of the original CNO isotopes.  

The evidence suggests that KP~Cyg probably belongs to the CWB variables rather 
than to the RR~Lyrae stars. Another RR~Lyrae star (Wallerstein et al. 2009), UY~CrB, 
seems to be very similar to KP~Cyg because it follows the same CNO trends shown 
in Fig.~2 and 3, and has a long period (0.929 days, Schmidt 2002). The two 
carbon cepheids, V553 Cen and RT TrA, first recognized by Lloyd Evans (1983)
and analysed by Wallerstein and Gonzalez (1996) and by Wallerstein, Matt and
Gonzalez (2000) have been included in Table 5 as well as Figures 2 and 3.
Their C and N excesses are similar to those of KP Cyg. In fact, there 
is no reason why variables should be classified according to whether their periods 
are greater or less than the rotation period of the Earth. Once the importance of the
so-called break at one day is disregarded, KP Cyg could also be a short period classical
 cepheid. Its low galactic latitude of 5 degrees certainly permits that.


\section{Conclusion}

KP~Cyg has an unusually long pulsational period ($P = 0.856$ days) for an RR~Lyrae star.  
If it is an RR~Lyrae star, KP~Cyg is expected to be a metal-poor star. However, the derived 
iron abundance completely rules out the possibility that KP~Cyg is a metal-poor star. Our 
analysis suggests that KP~Cyg is more likely a short-period CWB type star. Its low Galactic 
latitude of only 5 degrees is notable. The origin of the relatively metal-rich RR~Lyrae and 
CWB stars remains uncertain since they have not been related to another population such as a 
globular cluster of solar metallicity. It may be necessary to reach out to dE galaxies such 
as the companions to M31 to find  RR~Lyrae or CWB type stars of solar metallicity in a system 
that we understand better than our own complicated Galaxy. Perhaps the 30-m telescopes now being 
designed will be able to accomplish that.


\acknowledgements

SMA and VVK would like to express their gratitude to the Kenilworth Fund
of the New York Community Trust for the financial support of this study.
The individual financial support from Kenilworth Fund was made possible
through CRDF. SMA also thanks the Paris Observatory, Meudon, for its 
hospitality while this paper was in the final stages of preparation. 
 We thank Don York for the spectrum of Vesta, and Marta Mottini 
for reading the manuscript and making some good suggestions. We also thank
 the referee, George Preston, for his helpful comments. Much of the 
information about KP~Cyg was gathered with the help of SIMBAD.


\clearpage
\appendix
\begin{table*}
\tablenum{A1}
\begin{center}
\caption[]{Equivalent widths in the program spectra of KP Cyg}
\begin{tabular}{ccccccccc}
\hline
\hline
\multicolumn{3}{c}{}&
\multicolumn{1}{c}{$\phi=0.135$}&
\multicolumn{1}{c}{0.299}&
\multicolumn{1}{c}{0.430}&
\multicolumn{1}{c}{0.720}&
\multicolumn{1}{c}{0.800}&
\multicolumn{1}{c}{0.991}\\
\hline
\multicolumn{1}{c}{$\lambda$ (\AA)}&
\multicolumn{1}{c}{Ion$^a$}&
\multicolumn{1}{c}{$\log~gf$}&
\multicolumn{6}{c}{EW$^b$ (m\AA)}\\
\hline
\hline
\label{tableA1}
   ...   &   ...  &  ...  &   ... &  ... &   ...&   ...&  ...  &  ...\\ 
 6305.30 &  26.01 & -1.80 &    44 &    0 &   37 &    0 &     0 &    0\\
 6331.95 &  26.01 & -1.86 &    43 &   49 &   49 &    0 &     0 &   39\\
 6369.46 &  26.01 & -4.14 &    81 &   94 &   80 &   72 &    91 &    0\\
 6383.72 &  26.01 & -2.24 &    44 &   56 &   55 &   44 &    63 &   51\\
   ...   &   ...  &  ...  &   ... &  ... &   ...&   ...&  ...  &  ...\\ 

\hline
\hline
\end{tabular}
\\
\end{center}
$^a$ Code for Ions, e.g. 26.01 = FeII  \\
$^b$ EW = 0 means that the EW measurement of the line is not available. \\
\end{table*}


\begin{thebibliography}{}

 \bibitem[Andrievsky et al.(2001)]{}
Andrievsky, S. M., Kovtyukh, V. V., Korotin, S. A., Spite, M., \& Spite, F.\ 2001, A\&A, 367, 605 

   \bibitem[Andrievsky et al.(2007)]{}
Andrievsky, S. M., Spite, M., Korotin, S. A., Spite, F., Bonifacio, P., Cayrel, R., Hill, V., \& Fran\c cois P.\ 2007, A\&A, 464, 1081 

   \bibitem[Andrievsky et al.(2008)]{}
Andrievsky, S. M., Spite, M., Korotin, S. A., Spite, F., Bonifacio, P., Cayrel, R., Hill, V., \& Fran\c cois, P.\ 2008, A\&A, 481, 481  

   \bibitem[Andrievsky et al.(2009)]{}
Andrievsky, S. M., Spite, M., Korotin, S. A., Spite, F., Fran\c cois, P., Bonifacio, P., Cayrel, R., \& Hill, V.\ 2009, A\&A, 494, 1083 

   \bibitem[Andrievsky et al.(2010a)]{}
Andrievsky, S. M., Spite, M., Korotin, S. A., Spite, F., Bonifacio, P., Cayrel, R., Fran\c cois, P., \& Hill, V.\ 2010a, A\&A, 509, 88  

   \bibitem[Andrievsky et al.(2010b)]{}
Andrievsky, S. M., Spite, M., Korotin, S. A., Spite, F., Bonifacio, P., Fran\c cois, P., Cayrel, R., \& Hill, V.\ 2010b, A\&A, submitted  

   \bibitem[Carlsson(1986)]{}
Carlsson M., 1986, Uppsala Obs. Rep. 33

   \bibitem[Clement et al.(2001)]{}
Clement, C. M., et al.\ 2001, \aj, 122, 2587

   \bibitem[Clementini et al.(1995)]{}
Clementini, G., Carretta, E., Gratton, R., Merighi, R., Mould, J. R., \& McCarthy, J. K.\ 1995, AJ, 110, 2319

   \bibitem[Galazutdinov(1992)]{}
Galazutdinov, G. A.\ 1992, Preprint Special Astrophysical Observatory  RAS, 92 

   \bibitem[Gratton et al.(2007)]{}
Gratton, R. G., Lucatello, S., Bragaglia, A., Carretta, E., Cassisi, S., Momany, Y., Pancino, E., Valenti, E.,
Caloi, V., Claudi, R., D'Antona, F., Desidera, S., Fran\c cois, P., James, G., Moehler, S., Ortolani, S.,
Pasquini, L., Piotto, G., \& Recio-Blanco, A., 2007, A\&A, 464, 953

   \bibitem[Kinemuchi et al.(2009)]{}
Kinemuchi, K., Smith, H. A., Wozniak, P. R., \& McKay T.A.\ 2009, AJ, 132, 1202  

   \bibitem[Korotin(2009)]{}
Korotin, S. A. 2009, ARep, 53, 651  

    \bibitem[Korotin et al.(1999a)]{}
Korotin S.A., Andrievsky S.M., Luck R.E., 1999a, A\&A 351, 168

    \bibitem[Korotin et al.(1999b)]{}
Korotin S.A., Andrievsky S.M., Kostynchuk L.Yu., 1999b, Ap\&SS 260, 531

   \bibitem[Kovtyukh \& Andrievsky(1999)]{}
Kovtyukh, V. V., \& Andrievsky, S. M.\ 1999, A\&A, 351, 597            

   \bibitem[Kupka et al.(2000)]{}
Kupka, F., Ryabchikova, T. A., Piskunov, N. E. et al.\ 2000, Baltic Astron., 9, 

   \bibitem[Kurucz(1996)]{}
Kurucz, R.\ 1996, In Model Atmospheres and Spectrum Synthesis, ASPC, 108, 270

   \bibitem[Lloyd Evans(1983)]{}
Lloyd Evans, T.\ 1983, Observatory, 103, 276

   \bibitem[Loomis et al.(1988)]{}
Loomis, Ch., Schmidt, E. G., \& Simon, N. R.\ 1988, \mnras, 235, 1059  

   \bibitem[Lyubimkov et al.(2010)]{}
Lyubimkov, L. S., Lambert, D. L., Korotin, S. A., Poklad, D. B., Rachkovskaya, T. M., \& Rostopchin S.I.\ 2010, MNRAS, in press  

   \bibitem[Maas et al.(2007)]{}
Maas, T., Giridhar, S., \& Lambert, D. L.\ 2007, ApJ, 666, 378

    \bibitem[Mishenina et al.(2000)]{}
Mishenina, T. V., Korotin, S. A., Klochkova, V. G., \& Panchuk, V. E.\ 2000, A\&A, 353, 978 

   \bibitem[Preston(1959)]{}
Preston, G. W.\ 1959, ApJ 130, 507   
                          
   \bibitem[Pritzl et al.(2000)]{}
Pritzl, B., Smith, H. A., Catelan, M., \& Sweigart, A.V.\ 2000, ApJ, 530L, 41  

   \bibitem[Schmidt(1999)]{}
Schmidt, E. G.\ 2002, ApJ, 123, 965                          

   \bibitem[Smith(1995)]{}
Smith, H. A., RR Lyrae Stars, 1995, Cambridge University Press  

   \bibitem[Tsymbal(1996)]{}
Tsymbal, V. V.\ 1996, Model Atmospheres and Spectrum Synthesis, ed. S. J. Adelman,
F. Kupka \& W. W. Weiss (San Francisco),  ASP Conf. Ser., 108

    \bibitem[Vogt(1970)]{}
Vogt M., 1970, IBVS 468, 1

    \bibitem[Wallerstein \& Gonzalez(1996)]{}
Wallerstein, G., \& Gonzalez, G.\ 1996, \mnras, 282, 1236
    
    \bibitem[Wallerstein et al.(2000)]{}
Wallerstein, G., Matt, S., \& Gonzalez, G.\ 2000, \mnras, 311, 414

    \bibitem[Wallerstein et al.(2007)]{}
Wallerstein, G., Kovtyukh, V. V., \& Andrievsky, S. M. 2009, \aj, 133, 1373

    \bibitem[Wallerstein et al.(2009)]{}
Wallerstein, G., Kovtyukh, V. V., \& Andrievsky, S. M. 2009, \apj, 692L, 127

\end{thebibliography}
\end{document}